# Resistivity testing of palladium dilution limits in CoPd alloys for hydrogen storage

S. S. Das*, G. Kopnov and A. Gerber

School of Physics and Astronomy, Tel Aviv University,

Ramat Aviv, 69978 Tel Aviv, Israel

* Current address: School of Physical Science, National Institute of Science Education and Research Bhubaneswar, Jatni 752050, India.

Palladium satisfies most of the requirements for an effective hydrogen storage material with two major drawbacks: it has a relatively low gravimetric hydrogen density and is prohibitively expensive for large scale applications. Pd-based alloys should be considered as possible alternatives to a pure Pd. The question is how much one can dilute the Pd concentration in a variety of candidate materials while preserving hydrogen absorption capability. We demonstrate that the resistivity measurements of thin film alloy samples can be used for a qualitative high-throughput screening and study of the hydrogen absorbing properties over the entire range of palladium concentrations. Contrary to palladium-rich alloys where additional hydrogen scattering indicates a degree of hydrogen content, the diluted alloy films respond by a decrease of resistance due to their thickness expansion. Evidence of significant hydrogen absorption was found in thin CoPd films diluted to just 20% of Pd.

Corresponding author: A. Gerber, email: gerber@tauex.tau.ac.il

# Introduction.

A hydrogen-based economy is one of the favorite approaches towards maintaining the current technological progress and standards of living while lowering carbon dioxide emissions [1,2]. Unlike fossil fuels, the combustion of hydrogen does not generate carbon dioxide ($CO_2$), but water vapor only. The ultimate goals of a hydrogen-based economy include the production of hydrogen while generating minimal greenhouse gases, the development of efficient infrastructures for hydrogen storage and transport, and harnessing of its energy via fuel cells. Hydrogen storage and transportation are critical prerequisites to the realization of this vision and are among the most challenging issues to overcome. Hydrogen may be stored either under high pressure as a gas, cryogenically as a liquid, or in the solid state as a hydride [3 - 6]. The latter option offers the best theoretical combination of gravimetric and volumetric capacity, stability, and safety. The search for the optimal storage materials is far from being accomplished. The hydride materials studied so far either can work near room temperature but have low gravimetric density or have high gravimetric density but slow kinetics and can only release hydrogen at high temperatures.

Palladium has the potential to play a major role in practically every aspect of the envisioned hydrogen economy owing to its catalytic and hydrogen absorbing properties [7, 8]. Contrary to the majority of the known metal hydrides, palladium can absorb large quantities of hydrogen at room temperature and atmospheric pressure due to the dissociative adsorption of $H_2$ molecules. It satisfies most of the requirements for an effective hydrogen storage material with two major drawbacks: a relatively low gravimetric hydrogen density (Pd is a heavy metal) [9] and is prohibitively expensive for large scale applications. Pd-based alloys should be considered as possible replacements to a pure Pd for reducing the cost, increasing the gravimetric hydrogen density capacity, and probably, improving the diffusivity and kinetics of hydrogen loading and release.  Multiple studies were devoted to the thermodynamics and solubility of hydrogen in Pd alloys [10, 11], however, most of these works were limited to the alloys with a relatively low solute metal content [12 – 18]. Studies of alloys with strongly diluted Pd (up to about 50%) are focused mainly on the permeability of selected binary and ternary alloy membranes [19, 20].]

Pd forms alloys with a spectrum of metals, such as fcc Cu, Ag, Pt, Al,  Ni, and bcc Fe, V, Nb, Ta, Mo, and W. The number of potential materials for hydrogen storage is very large and to identify the optimal ones, multiple samples have to be produced and tested. For that, using combinatorial methods to study and develop new materials can be highly useful. High throughput (combinatorial) materials science methodology is a relatively new research paradigm that offers the promise of rapid and efficient materials screening, optimization, and discovery. High-throughput experiments are characterized by a synthesis of a "library" sample that contains the materials variation of interest (typically composition), and rapid and localized measurement schemes that result in massive data sets [21]. The goal of any hydrogen storage program is to manufacture materials on a kilogram scale. However, for rapid fabrication and screening purposes, it is more efficient to produce thin film samples with a wide range of compositional variations using standard film deposition techniques [22]. Traditional volumetric [23, 24] and gravimetric [25] measurements of hydrogen absorption can't be used with thin film samples due to a negligible amount of the active material. Thus, indirect screening techniques have to be used.

Hydrogen absorption in metals can be detected by monitoring characteristic changes in the crystallographic and electronic band structure of the host materials. Hydrogen occupies the interstitial sites in the fcc and bcc lattices and causes a large expansion of the host crystal [26]. The lattice expansion can be measured by various structure characterization tools, such as X-ray and neutron diffraction [27] but adaptation of these techniques for a rapid and massive screening is challenging. A technically simpler method of lattice expansion detection was suggested about a decade ago [28], in which extension of the hydrogenated film deposited on a bending cantilever was sensed by the deflected laser beams. The technique was suggested for a possible high-throughput screening and tested with several Mg-based samples. Another property affected by hydrogen absorption is electrical resistivity. Metallic Mg becomes insulating and optically transparent when transforms to magnesium hydride, the property used in the hydrogenography screening technique [29]. Pd, on the other hand, remains metallic in the hydride state, and its resistance can either increase or decrease in response to hydrogen absorption. An increase of the resistance is attributed to an enhancement of electron scattering in the hydride phase. The effect was extensively studied and used for constructing the hydrogen metal phase diagrams [30] and in hydrogen detection systems [31, 32]. Reduction of resistivity at hydrogen loading was observed in ultrathin and nano-gap films [33 - 36] and attributed to the lateral swelling of separated grains. As a result, the metallic clusters touch and create new conducting channels or shrink the inter-particle gaps, which results in an overall decrease in resistance. However, as demonstrated recently [37], the reduction of resistance in the hydrogenated state is not restricted to the nano-gap structures but is a general property of thin films grown on rigid substrates. Adhesion to the substrate surface prevents the lateral expansion of the hydrogenated films. Absence of the in-plane expansion is transformed to the out-of-plane extension enhanced by Poisson's effect. The elastic thickness expansion can reach 12.6% if the atomic ratio between Pd and the absorbed hydrogen is 1 [38]. Expanding thickness increases the current flow cross-section and reduces the measured electrical resistance. The change in the resistivity between the hydrogen-free and the hydrogenated states is a superposition of the enhanced hydride scattering $\Delta\rho_H$ and reduction of the resistance due to the thickness expansion, which can be presented as:

$$\Delta\rho = \frac{1}{1+\gamma}(\Delta\rho_H - \gamma\rho_0) \qquad (1)$$

where: $\rho_0$ is the resistivity before hydrogen loading, and $\gamma = \Delta t/t_0$ is the thickness expansion coefficient (strain). The negative thickness expansion term $(-\gamma\rho_0)$ is proportional to the initial resistivity and can be much larger than the positive hydride scattering term in the high resistivity samples. In bulk Pd and the low resistivity thick films, the scattering term is dominant. On the other hand, the thickness expansion term is expected to dominate in thin films with an enhanced surface scattering and alloys with high electrical resistivity. Indeed, the resistance response polarity of thin Pd, thick PdSiO$_2$ granular mixtures, and CoPd alloy films was found to change from positive to negative at the resistivity threshold of 50 μΩcm in 4% H$_2$ atmosphere [37]. As a rule, the resistivity of binary palladium alloys increases with the increasing concentration of the alloying component. For example, alloying of Pd with about 10 % at of V, Nb, or Mo increases the room temperature resistivity by the factor 4 - 8 [39]. The lattice expansion term in films with such resistivity is expected to be large and easily detected.

The resistance measurements allow not only to establish the very fact of hydrogen absorption but can also provide valuable information on the kinetics of the processes involved. As was demonstrated in Ref. [40], the time dependence of the hydride scattering term can differ significantly from the lattice response one. It was suggested that loading of the metal host with a large amount of hydrogen can create an out-of-equilibrium state. The stress built up by a rapid hydrogen absorption is released by a plastic lattice expansion. The latter can be much slower than the gas diffusion process. Thus, the lattice expansion associated with hydrogen absorption can be observable after a significant delay. The ability to differentiate between the hydrogen diffusion and the lattice response, between the reversible elastic and the irreversible plastic deformations is valuable for the basic understanding of the hydrogenation mechanisms.

Here, we demonstrate the resistive testing of hydrogen absorption in Co-Pd alloys over a full range of palladium concentrations. We'll show the evidence of hydrogen absorption in the alloys containing just 20% of Pd.

**Experimental.**

Co and Pd are soluble and form an equilibrium fcc solid solution phase at all compositions at room temperature [41]. Ferromagnetic CoPd alloys were used recently for magnetic and spintronics detection of hydrogen [42 – 45]. 15 nm thick polycrystalline $Co_xPd_{100-x}$ films were grown by rf-magnetron sputtering onto room-temperature glass substrates from two separate targets (99.99%) Co and Pd in the custom made deposition system.. Co atomic concentrations covered the entire range $0 \leq x \leq 100$ The base pressure prior to deposition was $5 \times 10^{-7}$ mbar. Sputtering was carried out at Ar-pressure of $5 \times 10^{-3}$ mbar. Composition of samples was controlled by rf-power of the respective sputtering sources. The typical deposition rate was 0.01 – 0.1 nm/sec. No post-deposition annealing was used. After the deposition the samples were mounted on the sample holder and connected with 20mm Si/Al wire using the wire-bonder. Resistance was measured following the Van der Pauw protocol. The measurement setup included the GMV 3473 electromagnet, Keithley 2400 source/meter, Keithley 2001 mutlimeter, and HP 3488A switch/control unit. The effect of hydrogen absorption was determined from measurements done in nitrogen and 4% $H_2/N_2$ mixture gas at ambient pressure and room temperature.

**Results and discussion.**

Accurate resistivity measurements of thick palladium films can be challenging because of buckling and cracks developing in the process of hydrogen absorption due to high compressive stress. An example of such damage in 20 nm thick Pd film is illustrated in Fig.1 (a,b). On the other hand, 15 nm thick and thinner films of Pd and CoPd alloys in the entire concentrations range are mechanically stable under repeating hydrogenation and dehydrogenation cycles, and no buckling or cracks were detected in any sample. Micrographs of the $Co_{30}Pd_{70}$ film surface before and after a hydrogenation – dehydrogenation cycle are shown in Figs 1c and 1d. 15 nm thick films are continuous and uniform. Films thinner than 10 nm have inhomogeneous meandric morphology with a continuous percolating metallic path across the sample. Films with thickness below 3 nm are discontinuous and not conducting. Thus, mechanically stable uniform 15 nm thick films were selected for this study.

Fig.2 presents the time dependent resistivity response to a sequence of hydrogen loading and unloading cycles (sequential exposure to 1 atm 4% $H_2/N_2$ mixture followed by $N_2$) of three $Co_xPd_{100-x}$ samples with x = 10 (a), x = 40 (b), and x = 80 (c). The starting resistivity grows gradually with increasing Co content from 35 μΩcm in $Co_{10}Pd_{90}$ to 110 μΩcm in $Co_{40}Pd_{60}$. The first exposure to hydrogen is different in samples with different Co concentration and resistivity. Resistance increases in the low resistivity $Co_{10}Pd_{90}$ (a) and decreases in samples with higher resistivity ($\rho_0 > 50$ μΩcm). The response to hydrogen removal is similar in samples (a) and (b): resistance decreases and saturates in $N_2$. The further response is reproducible and similar in all films: resistance increases when exposed to hydrogen and drops on its removal. The sequence of reproducible rapid increase/decrease responses to the loading/unloading of hydrogen superposed with an irreversible gradual reduction of resistivity can be interpreted as a superposition of reversible hydride formation-removal signals on the background of the irreversible thickness inflation. The response of low resistivity films is dominated by the hydride scattering term, while the lattice expansion is dominant in the high resistivity alloys. The reversible hydride formation-removal signal is negligibly small in $Co_{80}Pd_{20}$ sample and an irreversible reduction of resistivity is the only signal observed.

Different reversibility allows separating between the reversible hydrogen scattering and the irreversible lattice expansion contributions. The lattice expansion can be a long process depending not on the number of hydrogenation – dehydrogenation cycles but the duration of exposure to hydrogen. The process is irreversible in nitrogen – helium – nitrogen cycling. The magnitude of this expansion resistivity change was determined as the difference between the initial resistivity in nitrogen and the final one when the expansion process was over.

The scattering term is reproducible over a large number of hydrogenation – dehydrogenation cycles, as seen e.g. in Fig.2a. Moreover, the alloys kept their sorption – desorption capability for a long time, and the scattering response didn't change significantly when re-tested about a year after the fabrication. The magnitude of the effect decreases with increasing Co content. Fig. 3 presents the relative values of the scattering component, the thickness expansion term, and the total resistance change as a function of Co content. The electronic scattering term is about 25% in pure Pd, decreases with Co concentration, and becomes very small at x > 40%. On the other hand, the magnitude of the negative expansion term increases with Co concentration and is the largest at x = 70 - 80% before dropping to zero in pure Co. Co doesn't absorb hydrogen and no resistance changes were observed when a pure Co sample was exposed to hydrogen. Notably, the alloy samples with as much as 80% of Co demonstrated a strong lattice expansion response to hydrogen exposure, indicating hydrogen absorption.

The magnitude of the expansion response in strongly diluted alloys is very large however, the kinetics of the lattice expansion is very slow. Fig. 4 presents the effective half-time of the thickness expansion $T_{50}$, defined as the time at which the irreversible resistance reduction changed by half, as a function of Co concentration x. Hydrogen diffusion and the hydride formation accelerates with an increasing Co content down to a few sec in $Co_{40}Pd_{60}$ [40], which is quicker than in pure Pd. An enhanced diffusivity rate in these strongly diluted Pd-Co films is consistent with the conclusions of earlier studies of Pd-Au and Pd-Ag membranes [46, 47] and Au alloyed Pd surfaces [48]. The thickness expansion time scale is very different from the hydrogen diffusion one: 10 secs in the x = 10 sample up to $10^5$ sec in the x = 80% one. $T_{50}$ is fitted well as:

$T_{50} = T_0 e^{kx}$, with $T_0 = 13$ sec and $k = 0.1$ (red dashed line in Fig.4), i.e. it increases exponentially with Co concentration. The diffusion of hydrogen in and out of the material can be much quicker than the respective lattice response. The effect is interpreted as due to the creation of an non-equilibrium hydride state in which stress is built up rapidly with hydrogen absorption and is released by a slow plastic thickness growth which can be orders of magnitude slower than the gas diffusion time.

Resistivity measurements presented here provide qualitative evidence of hydrogen absorption in CoPd alloys but not quantitative information on hydrogen content. They can be used for high-throughput preliminary screening of alloys to evaluate the limits of palladium dilution. However, calibration of the resistance response by e.g. explicit volumetric measurements of macroscopic samples is required to evaluate the quantity of the stored hydrogen.

To summarize, we used the time dependent resistivity measurements to study the absorption capabilities of $Co_xPd_{100-x}$ films over a full range of Pd concentrations and found clear evidence of hydrogen absorption in strongly diluted (down to 20% of Pd) alloys. The dominant mechanism of the resistance response to hydrogen absorption in low resistivity Pd – rich alloys is scattering by interstitial hydrogen atoms. In high resistivity Pd – poor alloys the scattering mechanism becomes negligible. Instead, reduction of resistivity due to the lattice expansion is large and easily detectable. The lattice expansion and the respective reduction of resistivity are expected to be a general property of any hydrogen absorbing material. Resistivity measurements are technically simple and rapid. We propose to use the phenomenon and the method for high-throughput qualitative screening and study of the hydrogen-absorbing kinetics of Pd-based alloys. Our next step will be alloying of Pd with cheap and abundant Cu and Al.

- **Conclusions**
-

- Thin films (below 15 nm thick) of Pd and CoPd alloys are mechanically stable under repeating hydrogenation and dehydrogenation cycles, which allows performing reliable and reproducible resistivity measurements.
- The change in the resistivity between the hydrogen-free and the hydrogenated states is a superposition of the enhanced hydride scattering and reduction of the resistance due to the thickness expansion.
- The response of low resistivity Pd-rich films to hydrogenation is dominated by the hydride scattering term.
- Reduction of resistivity due to the lattice expansion is dominant in high resistivity Pd-poor alloys.

- Evidence of significant hydrogen absorption was found in CoPd alloys diluted to just 20% of Pd.
- The time scale of lattice expansion due to hydrogen absorption increases exponentially with Co concentration.
- Resistivity measurements can be used for qualitative high-throughput screening of Pd-based alloys over the entire concentration range.

## Acknowledgements.

The work was supported by the Israel Science Foundation grant No. 992/17.

# References.

22. A. Baldi and B. Dam, *Thin film metal hydrides for hydrogen storage applications*, J. Mater. Chem. 21, 4021 (2011).

23. J. J. Vajo, F. Mertens, C. C. Ahn, R. C. Bowman, B. Fultz, *Altering Hydrogen Storage Properties by Hydride Destabilization through Alloy Formation: LiH and MgH$_2$ Destabilized with Si*, J. Phys. Chem. B 108, 13977 (2004).

24. B. Bogdanovic, M. Schwickardi, *Ti-doped alkali metal aluminium hydrides as potential novel reversible hydrogen storage materials*, J. Alloys Compd. 253–254, 1 (1997).

25. J. Chen, N. Kuriyama, Q. Xu, H. T. Takeshita, T. Sakai, *Reversible Hydrogen Storage via Titanium-Catalyzed LiAlH$_4$ and Li$_3$AlH$_6$*, J.Phys.Chem. B 105, 11214 (2001).

26. Y. Fukai, *The Metal-Hydrogen System: Basic Bulk Properties,* Springer Series in Materials Science; Springer-Verlag: Berlin/Heidelberg, Germany, 2005; Volume 21.

27. S.J. Callori, C. Rehm, G.L. Causer, M. Kostylev and F. Klose, *Hydrogen Absorption in Metal Thin Films and Heterostructures Investigated in Situ with Neutron and X-ray Scattering*, Metals 6, 125 (2016).

28. A. Ludwig, J. Cao, A. Savan, and M. Ehmann, *High-throughput characterization of hydrogen storage materials using thin films on micromachined Si substrates*, J. Alloys Compd. 446, 516 (2007).

29. R. Gremaud, C.P. Broedersz, D.M. Borsa, A. Borgschulte, P. Mauron, H.Schreuders, J. H. Rector, B. Dam, and R. Griessen*, Hydrogenography: An Optical Combinatorial Method To Find New Light-Weight Hydrogen-Storage Materials,* Adv. Mater. 19, 2813 (2007).

30. F. D. Manchester, *Phase diagrams of binary hydrogen alloys* (ASM International, Materials Park, OH, 2000).

31. T. Hübert, L. Boon-Brett, G. Black, and U. Banach, *Hydrogen sensors – A review*, Sensors and Actuators B: Chemical 157, 329 (2011).

32. A. Mirzaei, H. R. Yousefi, F. Falsafi, M. Bonyani, J.-H. Lee, J.-H. Kim, H. W. Kim, and S. S. Kim, *An overview on how Pd on resistive-based nanomaterial gas sensors can enhance response toward hydrogen gas*, Int. J. Hydrogen Energy 44**,** 20552 (2019).
12

**Figure captions.**

Fig.1. Micrographs of 20 nm thick Pd (a, b), and 15 nm thick $Co_{30}Pd_{70}$ (c, d) film surfaces before and after the hydrogenation – dehydrogenation cycle respectively. Cracks are visible on the surface of hydrogenated Pd film.

Fig.2. Resistivity response to a sequence of hydrogenation - dehydrogenation cycles of three 15 nm thick $Co_xPd_{100-x}$ samples with x = 10 (a), x = 40 (b), and x = 80 (c).

Fig.3. The normalized resistance response to hydrogenation of CoPd films as a function of Co content x. Open dots indicate the electronic scattering term, solid dots – the thickness expansion term, and crosses – the total resistance change.

Fig.4. The half-time of the thickness expansion $T_{50}$ as a function of Co concentration x (open squares). Solid line (red) is a fit to: $T_{50} = T_0 e^{kx}$, with $T_0 = 13$ sec and $k = 0.1$.

**Figures**

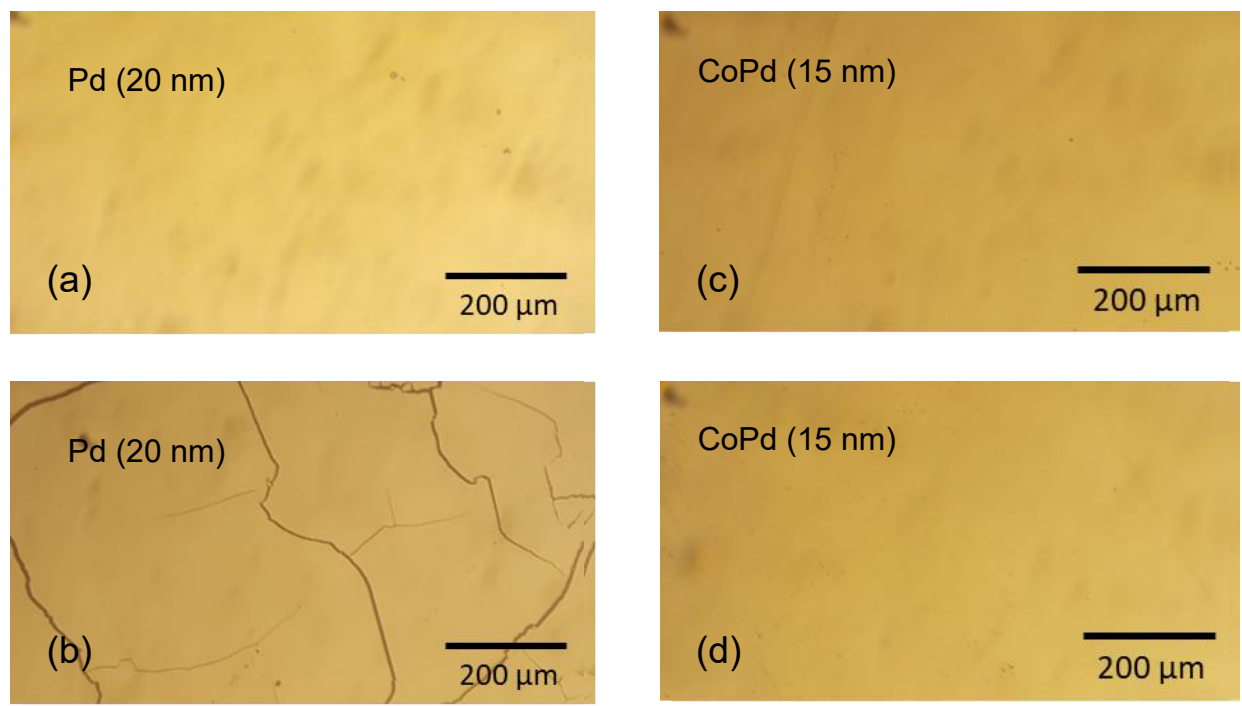

Fig. 1

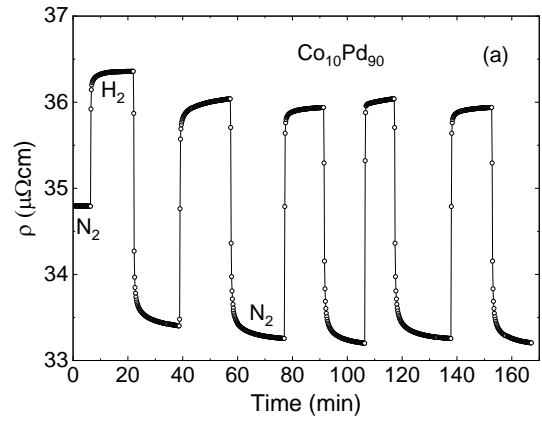

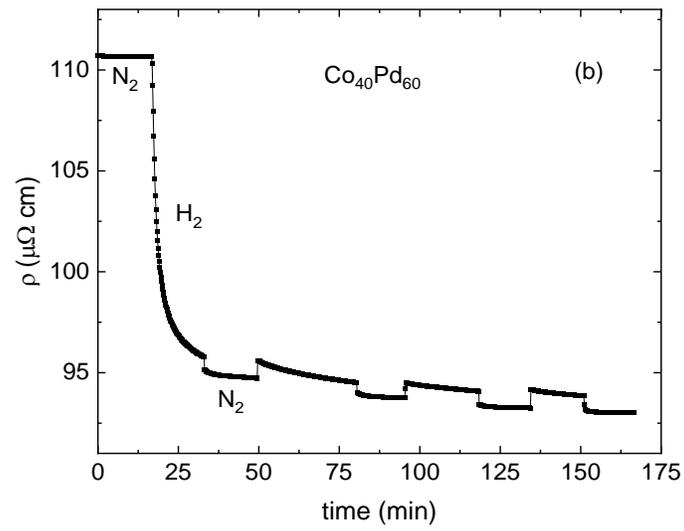

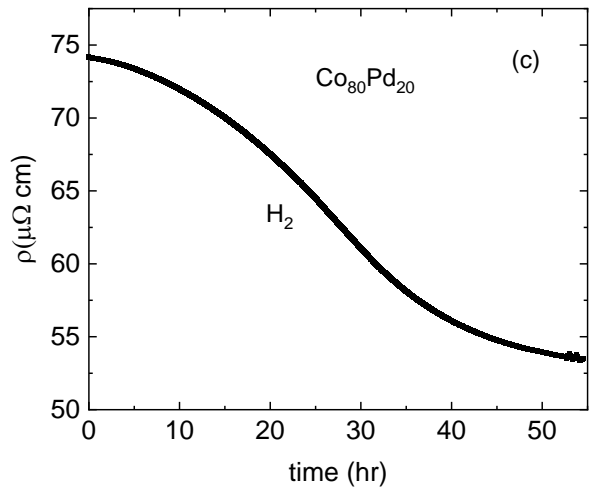

Fig. 2

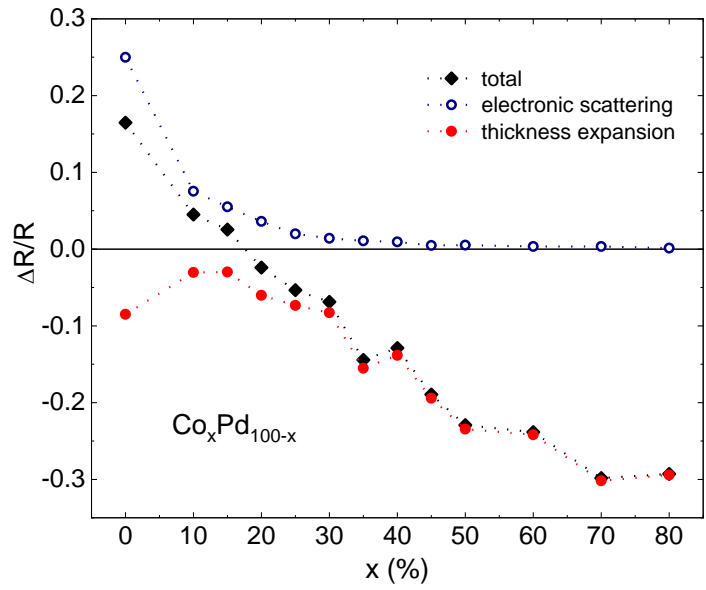

Fig. 3

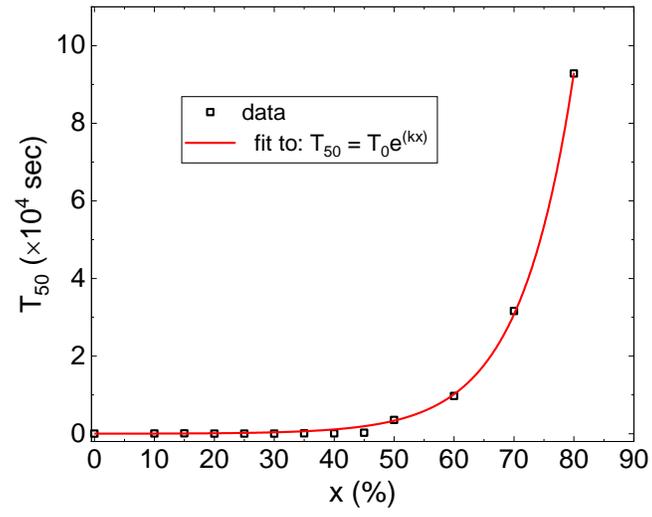

Fig. 4